\documentclass[]{spie}  
\usepackage[]{graphicx}
\usepackage{footnote} 
\makesavenoteenv{tabular}
\usepackage{wrapfig} 

\title{Commissioning ShARCS: the Shane Adaptive optics infraRed Camera-Spectrograph for the Lick Observatory 3-m telescope } 

\newcommand{\um}{$\mu$m}
\newcommand\arcsec{\mbox{$^{\prime\prime}$}}%

\author{Rosalie McGurk\supit{a}, Constance Rockosi\supit{a}, Donald Gavel\supit{b}, Renate Kupke\supit{b}, Michael Peck\supit{b}, Terry Pfister\supit{b}, Jim Ward\supit{b}, William Deich\supit{b}, John Gates\supit{b}, Elinor Gates\supit{b}, 
Barry Alcott\supit{b}, David Cowley\supit{b}, Daren Dillon\supit{b}, Kyle Lanclos\supit{b}, Dale Sandford\supit{b}, Mike Saylor\supit{b}, Srikar Srinath\supit{a}, Jason Weiss\supit{c}, and Andrew Norton\supit{b}.
\skiplinehalf
\supit{a}Astronomy Department and University of California Observatories-Lick Observatory, University of California, Santa Cruz, CA 95064, USA \\
\supit{b}University of California Observatories-Lick Observatory, University of California, Santa Cruz, CA 95064, USA \\
\supit{c}University of California Los Angeles Infrared Lab, University of California, Los Angeles, CA 90095, USA 
}

\authorinfo{Further author information: (Send correspondence to R.M.)\\R.M.: E-mail: rmcgurk@ucsc.edu, Telephone: 1 831 459 3068}

\begin{document} 
  \maketitle 

\begin{abstract}
We describe the design and first-light early science performance of the \textbf{Sh}ane \textbf{A}daptive optics infra\textbf{R}ed \textbf{C}amera-\textbf{S}pectrograph (ShARCS) on Lick Observatory's 3-m Shane telescope. Designed to work with the new ShaneAO adaptive optics system, ShARCS is capable of high-efficiency, diffraction-limited imaging and low-dispersion grism spectroscopy in J, H, and K-bands. ShARCS uses a HAWAII-2RG infrared detector, giving high quantum efficiency ($>$80\%) and Nyquist sampling the diffraction limit in all three wavelength bands. The ShARCS instrument is also equipped for linear polarimetry and is sensitive down to 650 nm to support future visible-light adaptive optics capability. We report on the early science data taken during commissioning.
\end{abstract}



\section{INTRODUCTION}
\label{sec:intro}  

Several near-infrared cameras have taken images corrected using the original laser guide star adaptive optics (AO) system that was installed on the Shane 3-m telescope at Lick Observatory in 1994\cite{Gilmore95,Bauman99}.  Prior to November 1998, the Lick InfraRed Camera (LIRC-2) was used as an interim science camera\cite{Gilmore95}; however, LIRC-2 was designed as a seeing-limited instrument, not optimized for AO performance.  In 1998, LIRC-2 was replaced by the InfraRed Camera for Adaptive optics at Lick Observatory (IRCAL)\cite{Lloyd00}. IRCAL was a 1.0-2.5 \um~camera and polarimeter optimized for use with the Lick laser guide star AO system. The Lick adaptive optics system was mounted at the f/17 Cassegrain focus of the Shane 3-m telescope, and delivered a corrected f/28.5 beam to IRCAL. Using diamond-turned gold-coated aluminum optics, IRCAL provided diffraction-limited imaging in the near-infrared (IR). IRCAL's detector was an anti-reflection coated Rockwell 256 $\times$~256 PICNIC array with 40 \um~pixels. IRCAL was Nyqusit-sampled for a corrected K-band beam, and offered a field of view of 20\arcsec. The Lick AO system generally produced Strehl ratios of 0.3 in K-band with laser guide star observing.

The new ShaneAO adaptive optics system\cite{Gavel11,Kupke12,Gavel14} is designed to produce diffraction-limited images in the J, H, and K-bands.  ShaneAO has two deformable mirrors in a woofer-tweeter wavefront correction system.  The first mirror, the ``woofer," is a high stroke, voice-coil actuated ALPAO deformable mirror; it provides tip-tilt and low-order wavefront correction.  The second mirror, the ``tweeter," is a high temporal and spatial bandwidth, low stroke Boston-Micromachines 32x32 MEMs deformable mirror, and provides high order corrections.  A high-sensitivity, high-bandwidth wavefront sensor camera with several Shack-Hartmann sampling modes is located after the second deformable mirror.  The current dye laser, in operation since 1996, will be replaced in the coming year with a fiber laser developed at Lawrence Livermore National Laboratories.  

The \textbf{Sh}ane \textbf{A}daptive optics infra\textbf{R}ed \textbf{C}amera-\textbf{S}pectrograph (ShARCS) is designed to work behind the ShaneAO system. ShARCS provides diffraction-limited and Nyquist-sampled images for J, H, and K-band beams over a 20\arcsec~field of view. Additionally, ShARCS is equipped for linear polarimetry and low resolution (R$\sim$500) long-slit spectroscopy. Section 2 discusses details of ShARCS's construction and testing.  Section 3 summarizes early science taken during commissioning.

\section{INSTRUMENT}

Much of the existing hardware from the IRCAL dewar was reused for ShARCS. The original liquid nitrogen cooled dewar, with a vapor-cooled radiation shield, houses ShARCS.  The addition of an ion pump has increased the hold time to at least 36 hours under normal conditions.  IRCAL's aperture wheel and two filter wheels are reused in ShARCS; located inside the cooled dewar, the three wheels are driven with new external servo motors via mechanical feedthroughs.  The aperture wheel contains two slits, a pinhole, an occulting finger, a half field used for polarimetry, and an open position, as well as two filter wheels. The filter wheels contain a variety of narrow and broad bandpass near-IR filters (see Table \ref{filtertable}), two low-resolution (R$\sim$500) grisms for spectroscopy, a pupil viewer to aid in alignment, and a Wollaston prism, used with a deployable waveplate in the adaptive optics relay, for polarimetry. For a summary of the various observing modes, please see Table \ref{modetable}.

The upgraded cryogenic and mechanical components of the instrument were manufactured by the University of Calfiornia Observatories technical shops in Santa Cruz, California. The original IRCAL components were made by IR Labs, Inc.

\begin{table}[t]  \begin{center}
\caption{ShARCS Filters \label{filtertable} }
\vspace{1em} 
\begin{tabular}{|l|c|c|c|c|} 
\hline
Filter Name & $\lambda_{den}$ 	& FWHM  					& Peak 			  & Normally \\
  	&(\um)\footnotemark[1]	& (\um)\footnotemark[1]  & Transmission (\%) & Available  \\
\hline
\hline
J	     						& 1.238 &	0.271 & 82 & Y \\
H							& 1.656 &	0.296 & 85 & Y \\
K							& 2.195 &	0.411 & 75 & Y \\
K$_{short}$				& 2.150 & 	0.320 & --\footnotemark[2] & Y \\
\hline
H$_2$ 1-0 S(1)			& 2.125 & 	0.020 & 80 & Y \\
BrGamma H I (n=7-4) 	& 2.167 & 	0.020 & 78 & Y \\
J CH$_4$ 				& 1.183 & 	0.040 & 86\footnotemark[3] & Y \\
K CH$_4$ 				& 2.356 & 	0.130 & 80 & Y \\
2.2/0.04					& 2.192 & 	0.047 & 79 & N  \\
H continuum				& 1.570 & 	0.020 & --\footnotemark[2] & N  \\
$[$Fe II$]$				& 1.644 & 	0.016 & --\footnotemark[2] & Y \\
K continuum				& 2.270 & 	0.020 & --\footnotemark[2] & N  \\
\hline 
\end{tabular} \\
\footnotemark[1]{Vacuum wavelengths} \\
\footnotemark[2]{Specified, not measured} \\
\footnotemark[3]{Unblocked, needs to be crossed with the J filter for blocking}\\
\end{center}
\end{table} 

\begin{table}[b]  
\begin{center}
\caption{ShARCS Observing Modes and Details\label{modetable}}
\begin{tabular}{|l|c|l|} 
\hline
Observing Mode  & Filters  				& Details \\
\hline
\hline
Imaging: 			& J, H, Ks, K bands	& 2, 3, 4, 4 pixels per lambda/D sampling, respectively \\
					& various narrow-band filters & All filters offer a field of view of 20\arcsec \\
\hline
Spectroscopy: 	& Hgrism, Kgrism		& R$\sim$500, slit 0.2\arcsec$\times$6.9\arcsec \\
\hline
Coronagraphy: 	& any 					& Finger diameter = 0.2\arcsec\\
\hline
Polarimetry: 		& Wollaston prism 	& Half Field aperture = 31.4\arcsec$\times$8.4\arcsec, for use with the \\
  					&						& externally mounted wave plate \\
\hline 
\end{tabular} 
\end{center}
\end{table} 

   \begin{figure}[ht!]
   \centering
   \begin{tabular}{c} 
   \includegraphics[width=0.8\columnwidth]{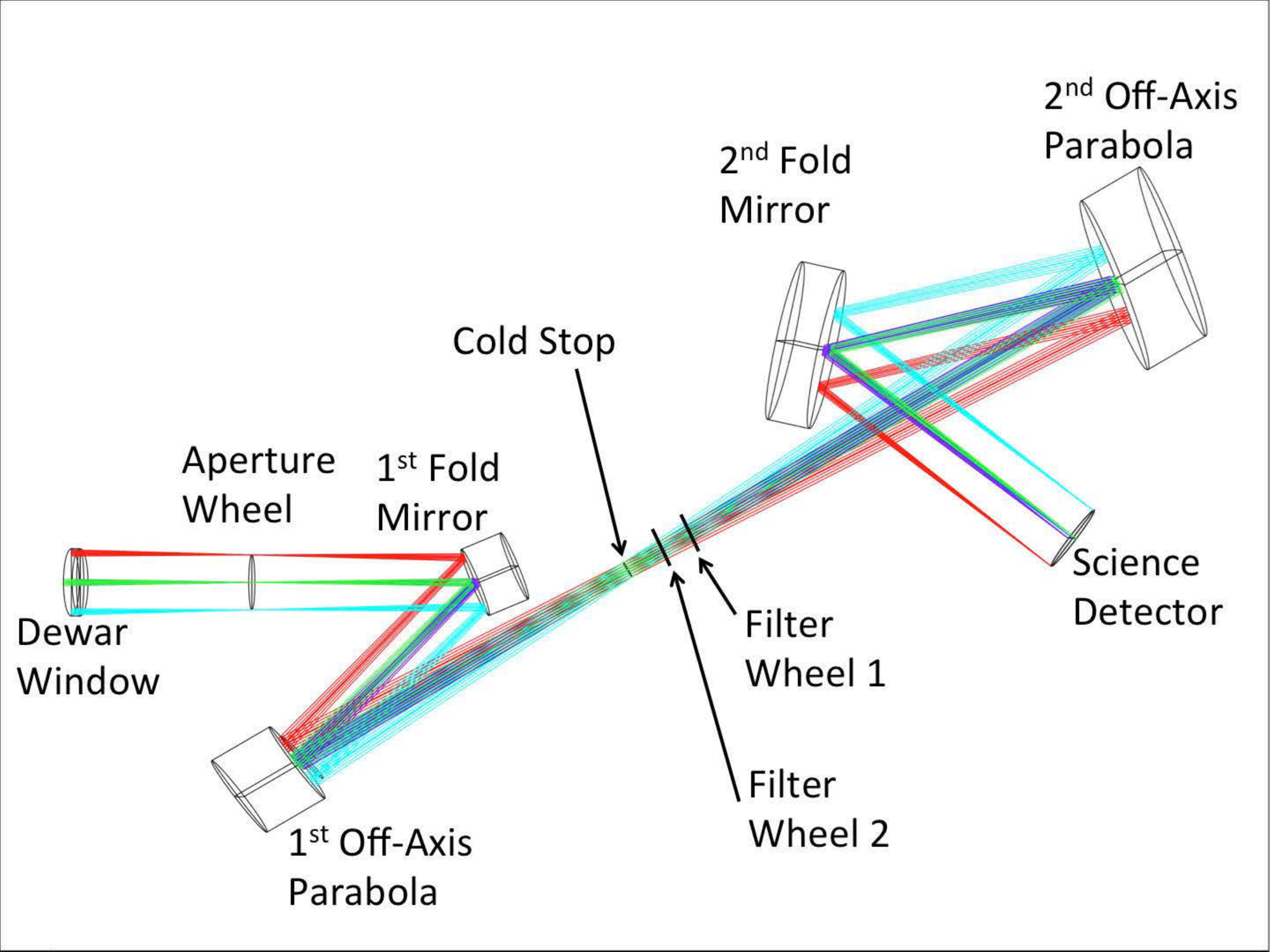} \\
   \includegraphics[width=0.8\columnwidth]{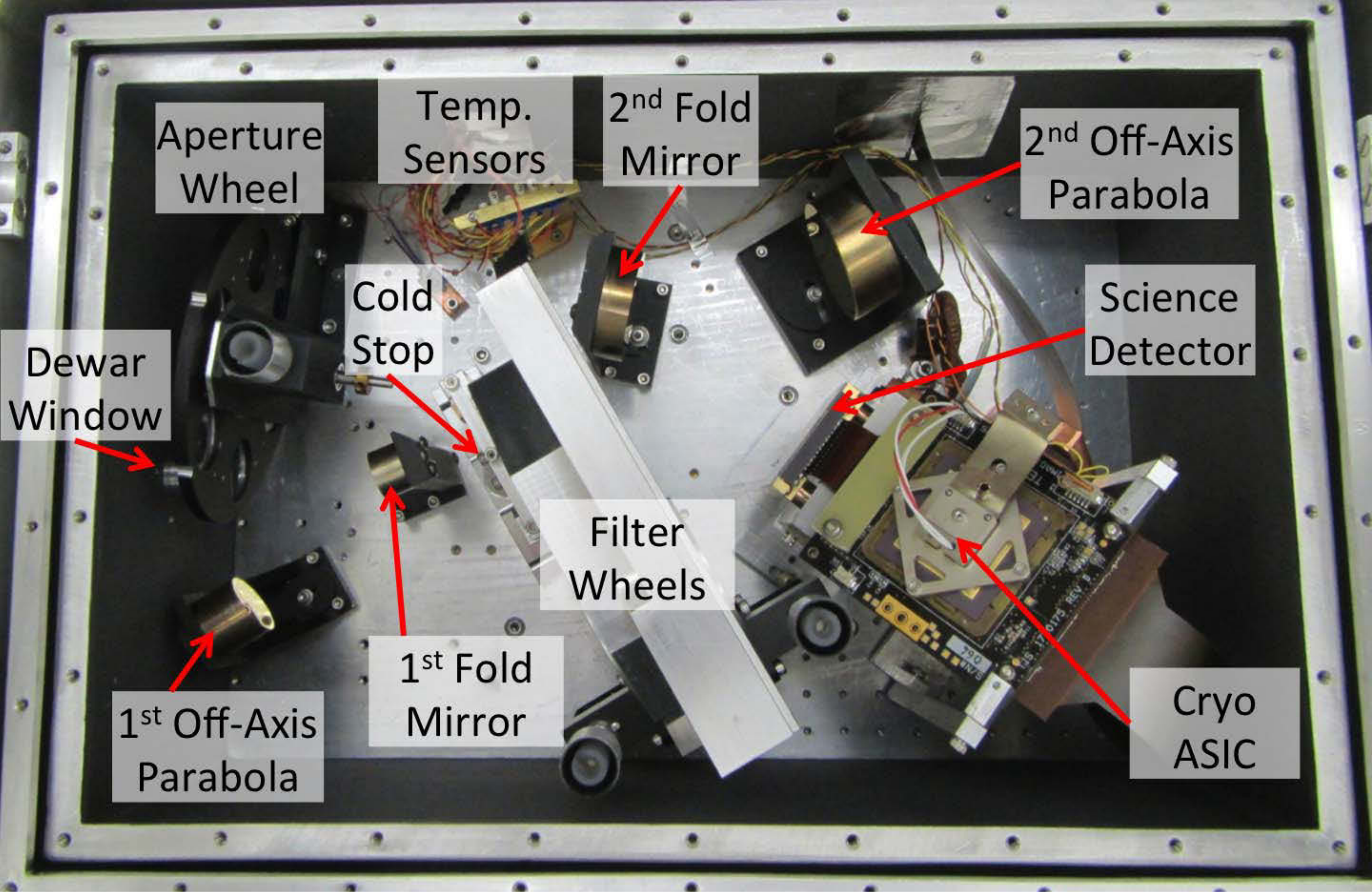}
   \end{tabular}
   \caption 
   { \label{opticaldesign} 
ShARCS Optical Design (top) and Optical Bench Image (bottom).  In both images, the light from the AO system enters on the left through a 0.85 inch diameter CaF$_2$ window. The converging f/28.5 beam forms a focal plane at the aperture wheel, is folded to the first off-axis parabola, and forms an image of the pupil on the 2.6 \% undersized cold stop.  The collimated beam passes through the two filter wheels, is focused by the second off-axis parabola, and folds onto the 5$^o$ tilted Teledyne H2RG detector read out by a cryogenic ASIC board.  Before taking this bottom photograph, the baffling around the beam both before and after the filter wheels has been temporarily removed for clarity.
} \vspace{-1em}
   \end{figure} 
\subsection{OPTICS}
The new optics used in ShARCS are an optimized pair of off-axis parabolas and two fold mirrors; all four optics are diamond-turned gold-coated aluminum.  As shown in Figure \ref{opticaldesign}, light from the AO system enters the dewar window and comes to a focus at the aperture wheel, deflects off the first fold mirror, is collimated by the first off-axis parabola, forms an image of the pupil at the cold stop (described below), passes through the two filter wheels, is focused by the second off-axis parabola, deflects off the fold mirror, and encounters the detector.

The Telescope Utilization Bin (TUB) is the rotatable mounting point for instruments using the Shane Telescopes's Cassegrain focus. Unlike the previous Lick adaptive optics system, ShaneAO is mounted to the TUB in a way that enables TUB rotation; this will allow the strategic orientation of the science object on the slit or occulting finger.  As the TUB rotates, the secondary support spiders will also appear to rotate as viewed in the pupil image at the science camera's cold stop and in the varying point spread function at the science detector. Since the previous AO system did not rotate, the cold stop included masking spiders to block the emission from the secondary support spiders.  ShaneAO does not mask the spiders since it is impractical to co-rotate the mask inside the dewar. Estimates of the thermal emissivity of the secondary support spiders indicated that they would not contribute significantly to the background seen at the science detector. This was confirmed by analysis of ShARCS commissioning images.

The ShARCS pupil image at the cold stop is 3.85 mm in diameter. To prevent the leakage of thermal background, the outer diameter of the new cold stop has been undersized by 100 \um, or 2.6\%, and the mask for the secondary obscuration has been oversized by 100 \um, or 2.6\%. The secondary support spiders are not masked.  The new cold stop is make from photomasked chrome with an optical density of 5 on a quartz substrate. The quartz has an anti-reflection coating on both sides.  Additionally, the cold stop has been relocated from downstream of the filter wheels to directly in front of the filter wheels, as seen in Figure \ref{opticaldesign}.  This will create a consistent point spread function for the entire spectrum diffracted by the grisms and both beams created by the Wollaston prism.

The aperture wheel in ShARCS contains a 100 \um~pinhole, a 100 \um~vertical slit for grism spectroscopy, a half-field mask for polarimetry, an open slot, a 100 \um~horizontal slit, and an occulting finger for coronagraphy.  The positioning accuracy in the aperture wheel has been improved in ShARCS so that the finger and slit now land within 2-3 pixels of their set positions.  Similar to IRCAL, ShARCS has a standard set of filters mounted, which can be swapped out by request.  The available filters are listed in Table \ref{filtertable}.  Filter wheel \#1, closest to the detector, contains a CaF K-band grism (R$\sim$500), an H-band Pupil Viewer, an open slot, and the following filters: Br Gamma, H$_2$ 1-0 S(1), K$_{short}$, H, and J. Filter wheel \#2, furthest from the detector and closest to the cold stop, contains a H-band grism (R$\sim$500), a solid blank for darks, an open slot, the Wollaston Prism, and the following filters: J CH$_4$, K CH$_4$, K, and $[$Fe II$]$.

   \begin{figure}[t]
   \begin{center}
   \includegraphics[height=7cm]{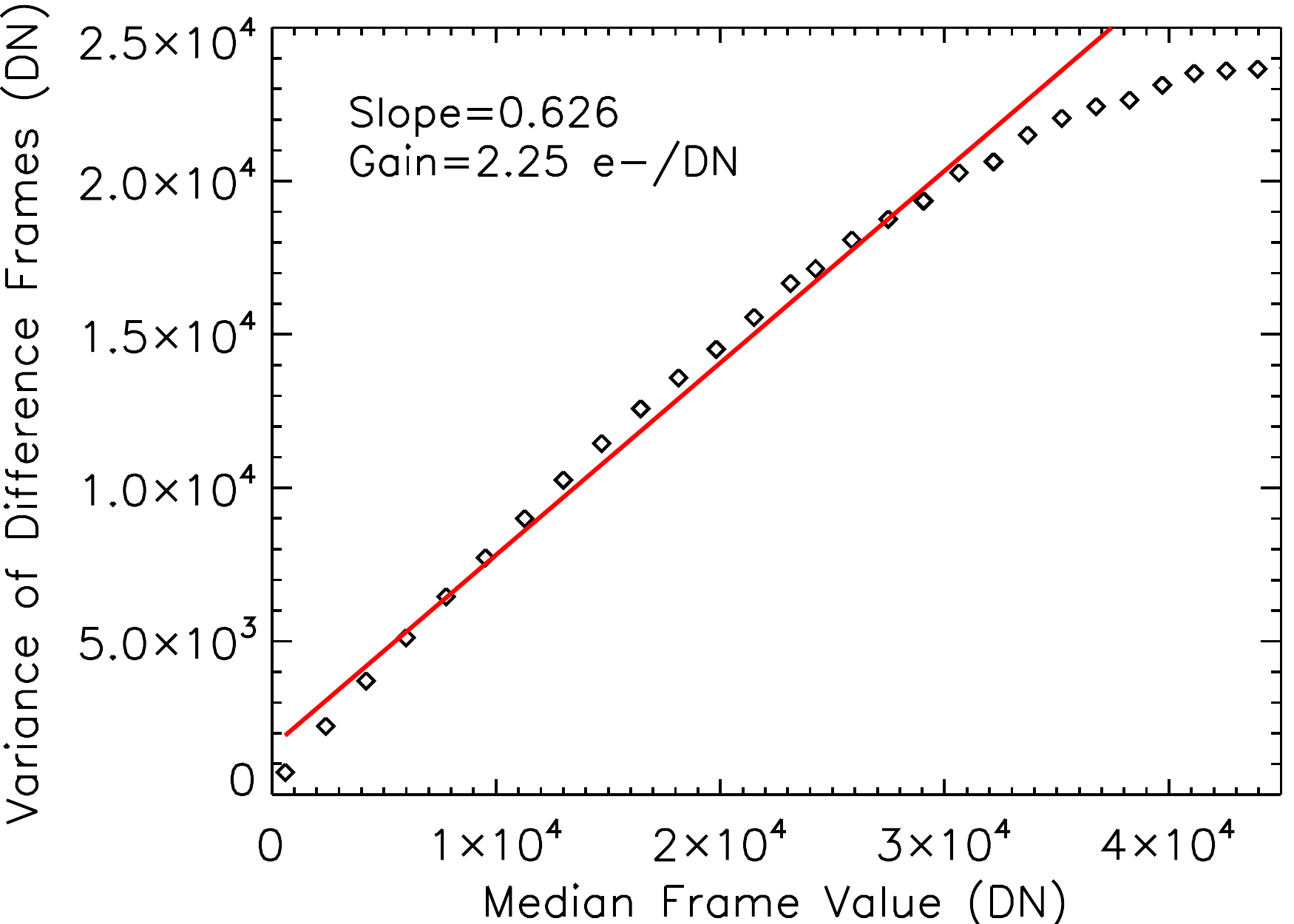}
   \end{center}
   \caption[example] 
   { \label{linearity} 
Linearity plot.  We took a series of image pairs of an evenly illuminated area of the dome with increasing exposure times. For this test, we select a 300 $\times$~300 square pixel box from which to measure our statistics (similar answers are found with a variety of box sizes and region choices) and remove the known bad pixels. The x-axis shows the 5-sigma clipped median value in DN for each frame, and the y-axis shows the 5-sigma clipped variance in the difference of the two frames in each image pair in DN. The overplotted line shows the fit to the linear portion of the data; the detector system is linear to about 35,000 DN, or about 82,000 e$^-$. From our measured slope of 0.626, we calculate the gain to be 2.25 e$^-$~DN$^{-1}$.}
   \end{figure} 

\subsection{DETECTOR DESCRIPTION}

The heart of the upgrade from IRCAL to ShARCS is replacing the science detector.  IRCAL used an anti-reflection coated Rockwell 256 $\times$~256 PICNIC array with 40 \um~pixels; it was sensitive from 0.85 -- 2.5 \um~with quantum efficiencies between 60 -- 65\%, and had a field of view of 19.5\arcsec.  The plate scale was 0.0780\arcsec ~pixel$^{-1}$~in the Right Ascension direction and 0.0754\arcsec~pixel$^{-1}$~in the Declination direction. IRCAL was diffraction-limited and Nyquist sampled in K-band.

ShARCS uses a Teledyne HAWAII-2RG (H2RG) HgCdTe engineering-grade near-infrared detector.  Out of the detector's 2048 $\times$~2048 square pixels, 1976 $\times$~1453 pixels (69.0\%) are operable and light sensitive, with 4 rows and columns of reference pixels along the sides of the detector.  We are using 600 $\times$~600 square pixels to cover the same $\sim$20\arcsec~field of view as IRCAL. If a higher resolution grism is installed in the future, the off-axis parabolic mirror that forms the final image of the science detector is over-sized to allow more of the detector's operable region to be illuminated for wider wavelength coverage. The decreased plate scale of 0.033\arcsec~pixel$^{-1}$~makes ShARCS diffraction-limited and Nyquist-sampled through the J, H, and K-bands.  The detector has a quantum efficiency of 82\%~over the wavelength range 0.6 -- 1.0 \um~and a quantum efficiency of 85\%~over the wavelength range 1.0 -- 2.5 \um.  

We chose to drive the H2RG with the SIDECAR application specific integrated circuit (ASIC) from Teledyne Imaging Sensors. The ASIC provides clocks and bias voltages to the detector and digitizes the detector outputs. As shown on the right in Figure \ref{opticaldesign}, the cryogenic ASIC is a small circuit board located inside the ShARCS dewar above the detector. A ribbon cable is potted through the dewar wall to connect the cryogenic ASIC to the Jade2 board, mounted on the outside dewar wall. The Jade2 provides the interface between the ASIC and USB 2.0.

\subsection{DETECTOR CHARACTERIZATION}

The detector can read out using 1, 4, or 32 output channels; 32 outputs gives the shortest base read time of 1.45 seconds, so this is the mode we generally utilize. We operate at gain setting 9, which translates to a preamp gain of 5.6. As shown in Figure \ref{linearity}, the detector system is linear to approximately 35,000 DN, or about 82,000 e$^-$.  The slope is related to the gain using: slope$=\sqrt{2}/$gain. From our measured slope of 0.626, we calculate the gain to be 2.25 e$^-$~DN$^{-1}$.

\begin{figure}[t]
   \begin{center}
   \includegraphics[height=10cm]{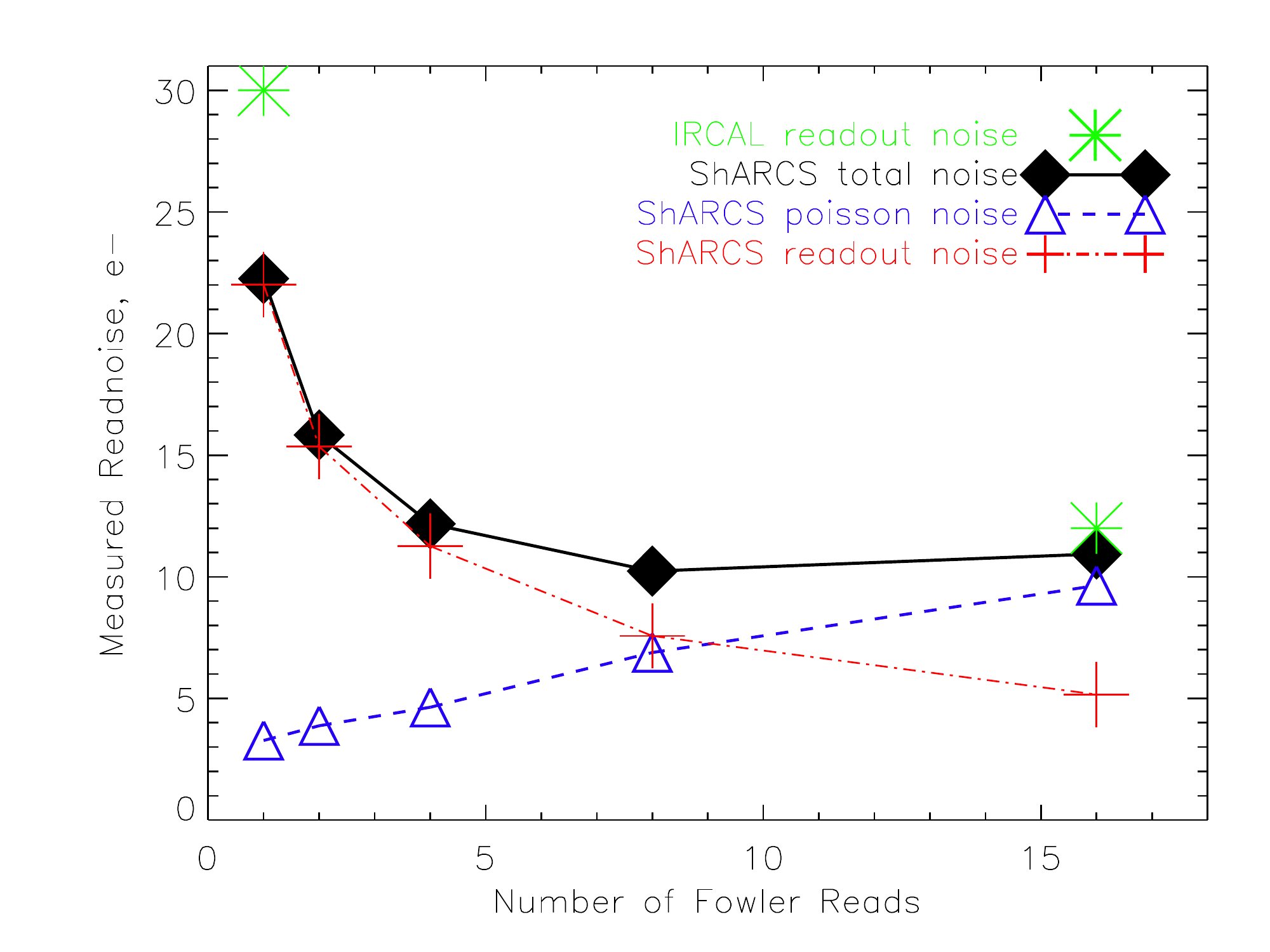}
   \end{center}
   \caption[example] 
   { \label{fowler} 
Fowler Pairs plot.  We took a series of dark images at the minimum exposure times possible for 1, 2, 4, 8, and 16 read pairs. The total noise averaged over the detector's full functional region is measured and plotted as filled black diamond points connected by a solid black line. The poisson noise from the measured charge accumulation during these increasing exposure times increases as shown in the blue dashed curve. The readout noise is calculated from the square root of the squared Poisson noise subtracted from the squared total noise; the readout noise is plotted as the red dot-dashed curve.  We calculate that ShARCS's H2RG and SIDECAR ASIC detector system has read noises of 22 e$^{-1}$~at Nreads=1 and 5 e$^{-1}$~at Nreads=16.  The two known points for IRCAL, Nreads=1 with a noise of 30 e$^{-}$~and Nreads=16 with a noise of 12 e$^-$, are overplotted as green asterisks for comparison. } 
   \end{figure}

Figure \ref{fowler} plots the measured total noise as a function of the number of Fowler reads using exposures of different durations, each just long enough to accommodate the set of multiple reads. The total noise, plotted as black diamonds connected by a black line, is measured from the variance in the difference of two frames with identical exposure times and numbers of reads, over the full functional detector area.  The measured charge accumulated in each frame increases as the number of reads, and thus minimum exposure time, increases.  This charge comes from dark current and additional background due to the incomplete baffling of the detector when these test data were taken. The Poisson noise contribution to the total measured noise in each difference frame is plotted as the blue dashed curve.  The read noise is calculated as  $\sqrt{ (\mbox{total noise})^2 - (\mbox{Poisson noise})^2}$, the square root of the squared Poisson noise subtracted from the squared total noise, and is plotted as the red dot-dashed curve.

IRCAL's PICNIC array had 30 e$^-$~of read noise with 1 read per pair (Correlated Double Sampling, CDS) and 12 e$^-$~of read noise with 16 reads per pair (Multi-Correlated Double Sampling, MCDS).  We find that our data on the ShARCS's H2RG and SIDECAR ASIC detector system in the final dewar is consistent with having 22 e$^-$~of read noise with 1 read per pair and 5 e$^-$~of read noise with 16 reads per pair.

\subsection{SOFTWARE}

A goal of the ShARCS software design was to use and build upon existing code created to run the H2RG in Keck Observatory's MOSFIRE instrument\cite{McLean10,McLean12}. This existing code was created at the University of California at Los Angeles Infrared Lab. The ShARCS computer architecture consists of two main computers: a detector control computer and an instrument host computer. The detector control computer's responsibility is to run the ShARCS ASIC Server and SIDECAR ASIC control software; this provides the low-level software interface to the Teledyne controller. This computer runs Microsoft's Windows XP Professional and is mounted in the instrument and ShaneAO electronics cabinet. The instrument host computer interfaces with the detector control server  and connects to the data storage archive. This system runs CentOS Linux 6.3 and lives in the Shane 3-m dome computer room. Additional top-level software controlled by observers uses the interface provided by the Linux computer to configure the instrument and take data.

ShARCS detector software employs a client-server architecture which facilitates distributed processing and remote observing. The MOSFIRE Detector Server (MDS) was implemented as a shareable library driven by the Keck Observatory standard application \textit{rpc\_server} which used Sun remote procedure calls (RPC) to implement the standard Keck Task Library (KTL) communications\cite{Kwok02}. When we received the software, \textit{rpc\_server} had Solaris dependencies that prevented it from running on our Linux platform, and we therefore created a substitute application, \textit{mr\_server}, that would run the MDS library without modification, but without using RPC-based communications. Other software was unaffected by the server-side change, and did not require any modification. Like the Keck standard, simple client commands, such as ``show" and ``modify", facilitate easy access to and the scripting of the server's keywords. By using existing KTL interfaces to other popular languages such as Java, IDL and Tcl, a diverse range of user interfaces can be implemented.  

From the observer's top-level view, the ShARCS software is broken into several different sub-units.  \textit{sharcs\_fe} is the graphical user interface (GUI) written in Java that interfaces with \textit{mr\_server}, configures the detector, takes exposures, and manages scripts that nod the telescope, take exposures, and control the Adaptive Optics Loops.  \textit{sharcswheels\_gui} is a Python GUI that allows observers to move the filter and aperture wheels.  \textit{sharcsdisplay} is an Tcl and IDL-based imaged display widget that allows the observer to manipulated the image using either the GUI or regular IDL commands.  \textit{eventsounds} is a program that allows the observer to pick which sounds signal different events in \textit{sharcs\_fe}, such as the end of exposures and the end of a script. 

   \begin{figure}[t]
   \begin{center}
   \includegraphics[width=6cm]{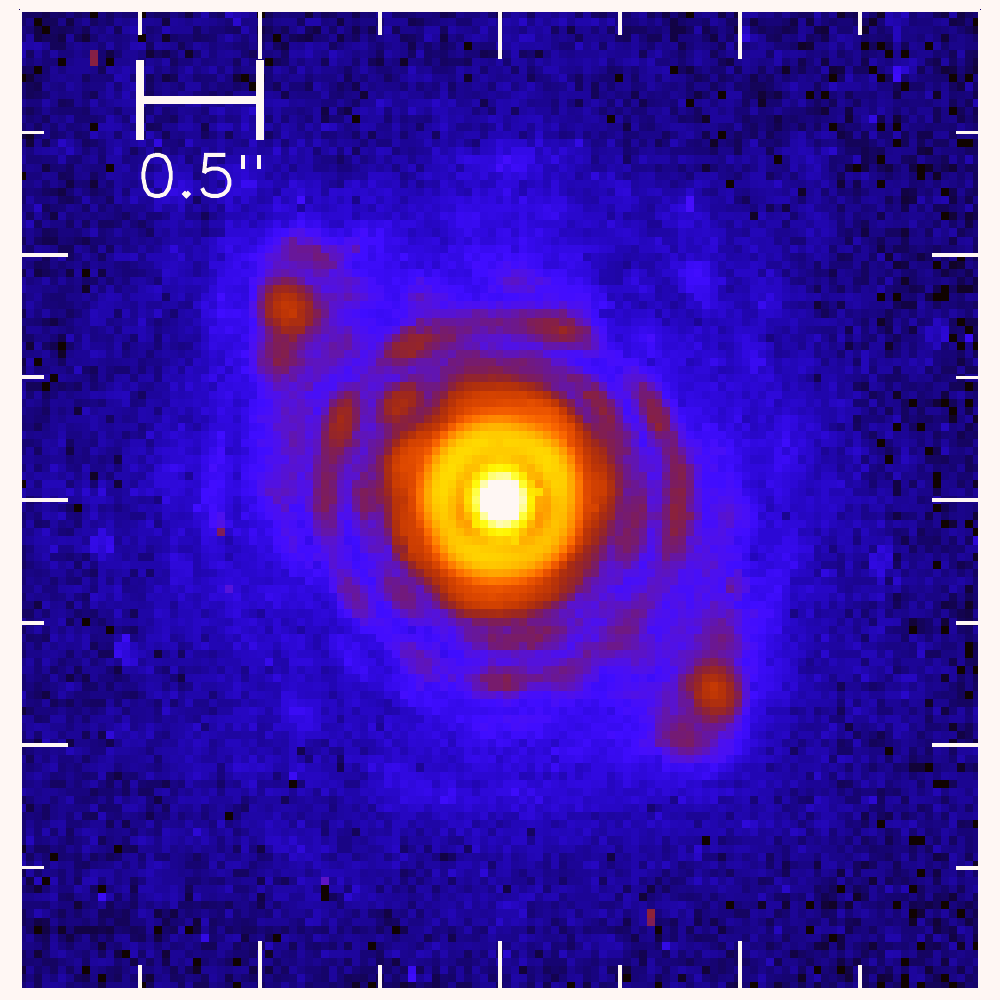}
   \end{center}
   \caption[example] 
   { \label{firstlight} 
First light image of Phi Gem, taken in the first hour on-sky. The narrow Br$\gamma$~filter with $\lambda_{cen}$=2.167 \um~is used. Image is displayed on a logarithmic scale. Up to 7 Airy rings are visible; using the \textit{sharcsstrehl} IDL tool, we measure a strehl of 0.52. We measure a FWHM of 0.188\arcsec, while the diffraction limit in Br$\gamma$~is 0.149\arcsec.}
   \end{figure} 

\section{PERFORMANCE}

\begin{figure}[t]
   \begin{center}
   \includegraphics[width=\textwidth]{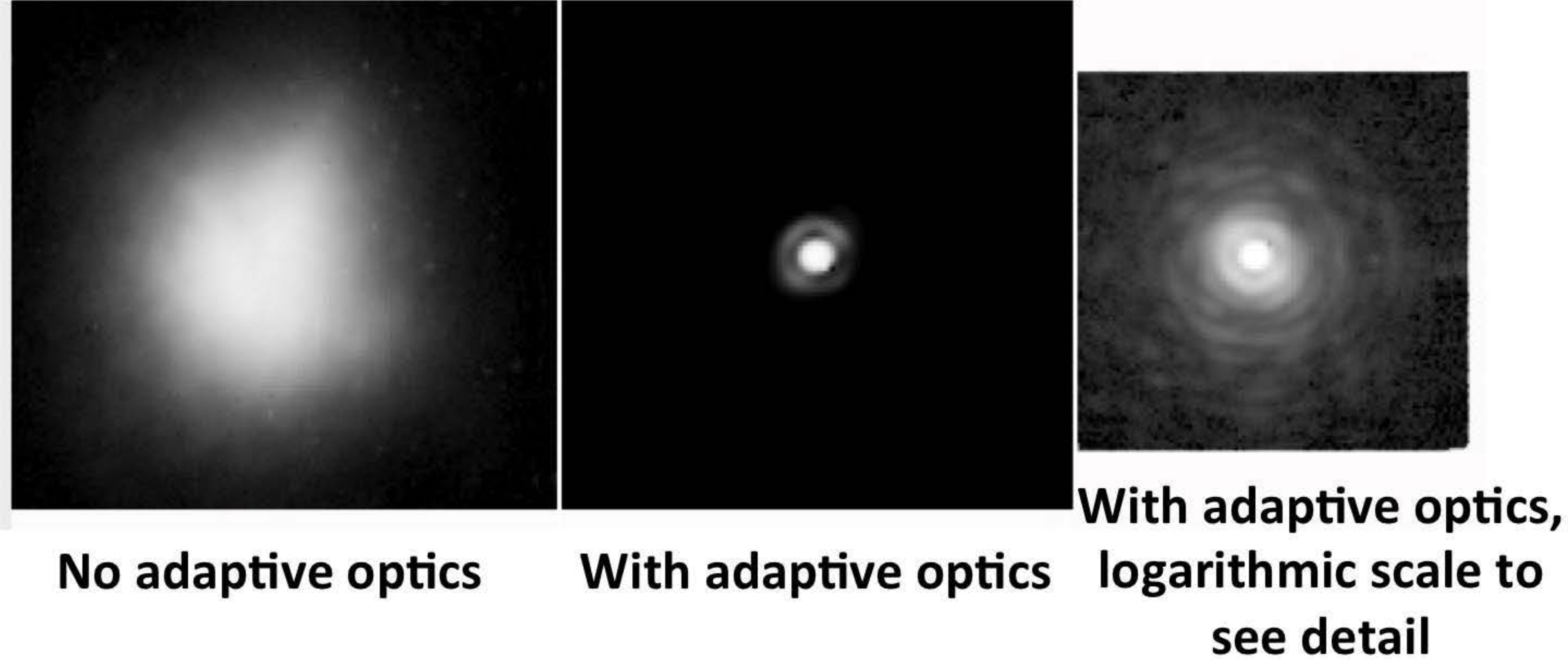}
   \end{center}
   \caption[example] 
   { \label{aoonoff} 
ShARCS images of Phi Gem, taken in the narrow Br$\gamma$~filter with $\lambda_{cen}$=2.167 \um. \textit{(left)} Image with the adaptive optics not correcting the atmosphere's turbulence. \textit{(middle)} Image with the adaptive optics system on. \textit{(right)}  The Airy rings can be seen more clearly when shown on a logarithmic scale.  Approximately four full rings and two more partial rings can be seen}
   \end{figure}

Pictures and discussion go here.  First light occurred on 2014 April 12 during the first commissioning run.  An hour after we got ShaneAO on-sky, we took our first light image of Phi Gem, a V=5.0 star, in narrow Br$\gamma$~filter with $\lambda_{cen}$=2.167 \um. The first light image is shown in Figure \ref{firstlight}, and, taken after tuning the AO system, a comparison pair of images with the AO system off and on are shown in Figure \ref{aoonoff}. Up to 7 Airy rings are visible. Using the \textit{sharcsstrehl} IDL tool, we measure a strehl of 0.52. We measure a FWHM of 0.188\arcsec, while the diffraction limit in Br$\gamma$~is 0.149\arcsec.
Srinath et al. 2014 compare measured results to predicted signal-to-noise ratio and magnitude limits from modeling the emissivity and throughput of ShaneAO and ShARCS\cite{Srinath14}.

\begin{figure}[h!]
   \begin{center}
   \includegraphics[width=\textwidth]{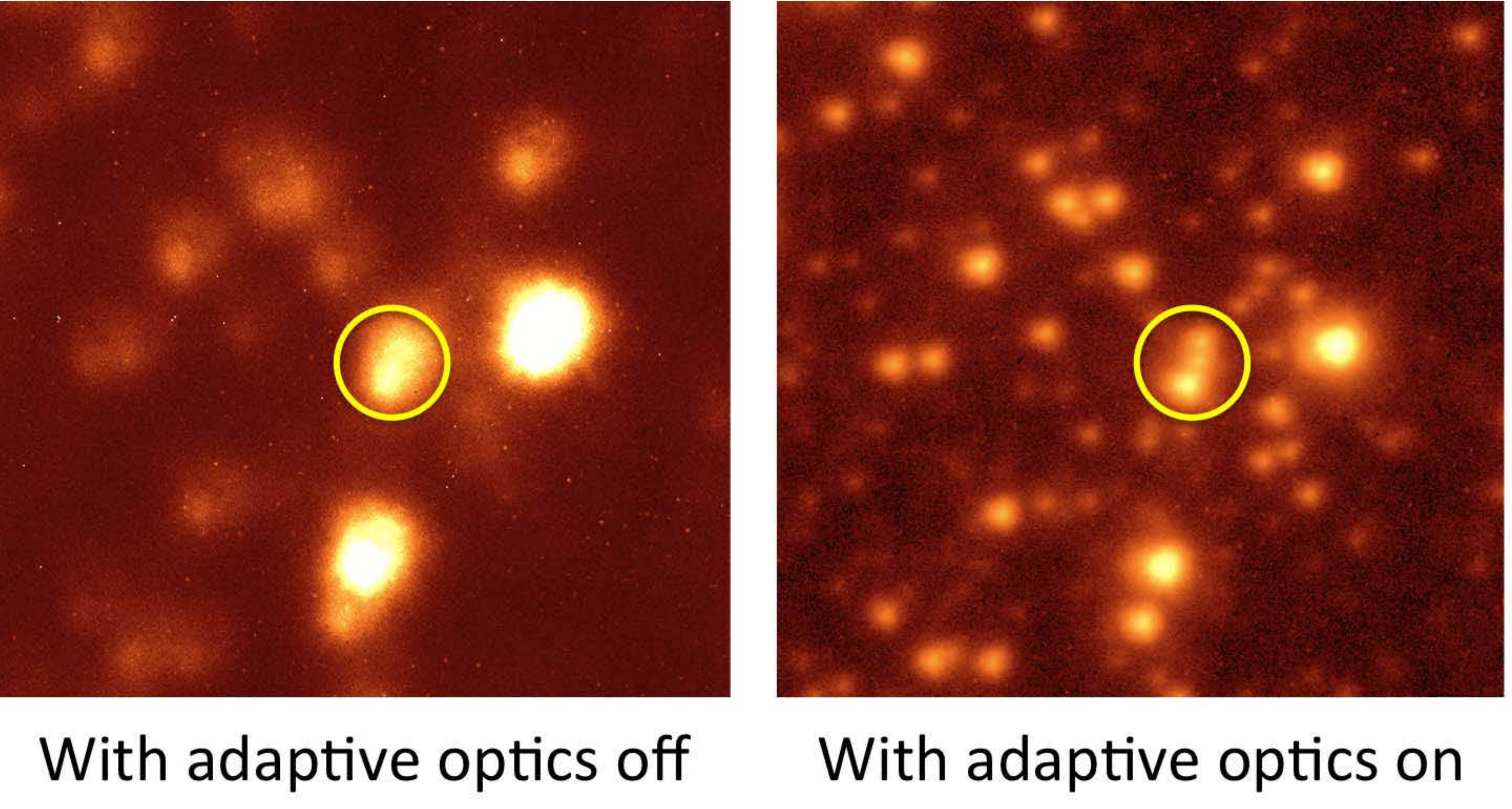}
   \end{center}
   \caption[example] 
   { \label{M92onoff} 
ShARCS image of M92, taken in the broad Ks filter with $\lambda_{cen}$=2.150 \um. \textit{(left)} Image with the adaptive optics not correcting the atmosphere's turbulence. \textit{(right)} Image with the adaptive optics system on. The three stars in the overplotted yellow circle are 0.42\arcsec~and 0.58\arcsec~apart.  For reference, the width of the stars in the ``AO off" image is 1\arcsec.}
   \end{figure}

Figure \ref{M92onoff} shows a 20\arcsec$\times$20\arcsec~portion of the globular cluster M92. This image shows the spatial resolution of the new ShaneAO system.  The frame on the left shows the data with no AO correction.  The panel on the right shows the same field, over the same exposure time, with the new AO system turned on.  The three stars in the circle are 0.42\arcsec~and 0.58\arcsec~apart.  For reference, the width of the stars in the ``AO off" image is 1\arcsec. This image was taken the second good night on sky.  ShaneAO can do better:  image sharpening can remove static aberrations from the AO system and telescope.  M92 is a famous and well-studied cluster, which is why we observed it for commissioning. But the reference  star used to compute the AO correction is faint, and we can optimize for better correction with faint reference stars.
We will do that during engineering runs this spring.

\acknowledgments     
 
ShARCS and ShaneAO are made possible through an National Science Foundation Major Research Instrumentation grant, \#0923585.  



\begin{thebibliography}{10}

\bibitem{Gilmore95}
{Gilmore}, K., {Rank}, D., and {Temi}, P., ``{The Lick Observatory Two Micron
  Camera},'' in [{\em New Developments in Array Technology and
  Applications}{\nolinebreak\hspace{0.1em}]},  {Philip}, A.~G.~D., {Janes}, K.,
  and {Upgren}, A.~R., eds., {\em IAU Symposium} {\bf 167},  79 (1995).

\bibitem{Bauman99}
{Bauman}, B.~J., {Gavel}, D.~T., {Waltjen}, K.~E., {Freeze}, G.~J., {Keahi},
  K.~A., {Kuklo}, T.~C., {Lopes}, S.~K., {Newman}, M.~J., and {Olivier}, S.~S.,
  ``{New optical design of adaptive optics system at Lick Observatory},'' in
  [{\em Adaptive Optics Systems and Technology}{\nolinebreak\hspace{0.1em}]},
  {Tyson}, R.~K. and {Fugate}, R.~Q., eds., {\em Proc. SPIE} {\bf 3762},
  194--200 (Sept. 1999).

\bibitem{Lloyd00}
Hanson, K.~M., ``Ircal: The infrared camera for adaptive optics at lick
  observatory,'' in [{\em Optical and IR Telescope Instrumentation and
  Detectors}{\nolinebreak\hspace{0.1em}]},  Iye, M. and Moorwood, A. F.~M.,
  eds., {\em Proc. SPIE} {\bf 4008},  814--821 (2000).

\bibitem{Gavel11}
Gavel, D.~T., ``Development of an enhanced adaptive optics system for the lick
  observatory shane 3-meter telescope,'' in [{\em MEMS Adaptive Optics
  V}{\nolinebreak\hspace{0.1em}]},  Olivier, S.~S., Bifano, T.~G., and Kubby,
  J.~A., eds., {\em Proc. SPIE} {\bf 7931},  793103--1--793103--8 (2011).

\bibitem{Kupke12}
{Kupke}, R., {Gavel}, D., {Roskosi}, C., {Cabak}, G., {Cowley}, D., {Dillon},
  D., {Gates}, E.~L., {McGurk}, R., {Norton}, A., {Peck}, M., {Ratliff}, C.,
  and {Reinig}, M., ``Ircal: The infrared camera for adaptive optics at lick
  observatory,'' in [{\em Adaptive Optics Systems
  III}{\nolinebreak\hspace{0.1em}]},  Ellerbroek, B.~L., Marchetti, E., and
  V\'{e}ran, J.-P., eds., {\em Proc. SPIE} {\bf 8447},  7pp. (2012).

\bibitem{Gavel14}
{Gavel}, D., {Kupke}, R., {Dillon}, D., {Norton}, A.~P., {Ratliff}, C.~T.,
  {Cabak}, G., {Phillips}, A.~C., {Rockosi}, C., {McGurk}, R., {Srinath}, S.,
  {Peck}, M., {Saylor}, M., {Ward}, J., and {Deich}, W.~T., ``{ShaneAO: wide
  science spectrum adaptive optics system for the Lick Observatory},'' in [{\em
  Status of Current AO Instrument Projects I}{\nolinebreak\hspace{0.1em}]},
  {\em Proc. SPIE} (2014).

\bibitem{McLean10}
{McLean}, I.~S., {Steidel}, C.~C., {Epps}, H., {Matthews}, K., {Adkins}, S.,
  {Konidaris}, N., {Weber}, B., {Aliado}, T., {Brims}, G., {Canfield}, J.,
  {Cromer}, J., {Fucik}, J., {Kulas}, K., {Mace}, G., {Magnone}, K.,
  {Rodriguez}, H., {Wang}, E., and {Weiss}, J., ``{Design and development of
  MOSFIRE: the multi-object spectrometer for infrared exploration at the Keck
  Observatory},'' {\em Proc. SPIE} {\bf 7735} (July 2010).

\bibitem{McLean12}
{McLean}, I.~S., {Steidel}, C.~C., {Epps}, H.~W., {Konidaris}, N., {Matthews},
  K.~Y., {Adkins}, S., {Aliado}, T., {Brims}, G., {Canfield}, J.~M., {Cromer},
  J.~L., {Fucik}, J., {Kulas}, K., {Mace}, G., {Magnone}, K., {Rodriguez}, H.,
  {Rudie}, G., {Trainor}, R., {Wang}, E., {Weber}, B., and {Weiss}, J.,
  ``{MOSFIRE, the multi-object spectrometer for infra-red exploration at the
  Keck Observatory},'' {\em Proc. SPIE} {\bf 8446} (Sept. 2012).

\bibitem{Kwok02}
{Kwok}, S.~H. and {Cohen}, R.~W., ``{Keywords revisited},'' in [{\em Advanced
  Global Communications Technologies for Astronomy
  II}{\nolinebreak\hspace{0.1em}]},  {Kibrick}, R.~I., ed., {\em Proc. SPIE}
  {\bf 4845},  194--200 (2002).

\bibitem{Srinath14}
{Srinath}, S., {Rockosi}, C., {McGurk}, R., {Kupke}, R., {Gavel}, D., {Gates},
  E., {Peck}, M., and {Dillon}, D., ``{Swimming with ShARCS: comparison of
  on-sky sensitivity with model predictions for ShaneAO on the Lick Observatory
  3-meter telescope},'' in [{\em Status of Current AO Instrument
  Projects}{\nolinebreak\hspace{0.1em}]},  {\em Proc. SPIE} (2014).

\end{thebibliography}

\end{document}